\documentclass[a4paper,fleqn,usenatbib]{mnras}

\usepackage{newtxtext,newtxmath}
\usepackage[T1]{fontenc}
\usepackage{ae,aecompl}
\usepackage{graphicx}	
\usepackage{amsmath}	
\usepackage{amssymb}	
\usepackage{ulem}

\title{Broadband X-ray characteristics of the transient pulsar GRO J2058+42}

\author[S. Kabiraj and B. Paul]{Sanhita Kabiraj$^{1,2}$\thanks{Contact e-mail: \href{mailto:sanhita@rri.res.in, 
sanhita@iisc.ac.in}
{sanhita@rri.res.in, sanhita@iisc.ac.in}} and Biswajit Paul$^{1}$\\
$^{1}$Raman Research Institute, Sadashivnagar, Bangalore-560080, India\\
$^{2}$Joint Astronomy Programme, Indian Institute of Science, Bangalore-560012, India\\}

\date{Accepted 2020 July 04. Received 2020 June 12; in original form 2019 August 13}

\pubyear{2015}

\begin{document}
\label{firstpage}
\pagerange{\pageref{firstpage}--\pageref{lastpage}}
\maketitle

\begin{abstract}
The Be X-ray binary GRO J2058+42 recently went through a Type-II outburst during March-April 2019 lasting for about 50 days. This outburst was detected with the operating all sky X-ray monitors like the  {\it{Fermi}}-GBM, {\it{Swift}}-BAT and {\it{MAXI}}-GSC. Two {\it{NuSTAR}} observations were also made, one during the rise and other during the decay of the outburst. It gave us the unique opportunity to analyze the broadband characteristics of the pulsar for the first time and accretion torque characteristics of the pulsar over a range of X-ray luminosity. The pulse profiles are strongly energy dependent, with at least four different pulse components at low energy (< 20 keV) which evolves to a single-peaked profile at high energy (> 30 keV). In each of the narrow energy bands, the pulse profiles are nearly identical in the two {\it{NuSTAR}} observations. The spectra from both the observations are fitted well to a power-law with a Fermi-Dirac type high energy cutoff. We ruled out presence of a cyclotron line in the pulse phase averaged X-ray spectrum in the {\it{NuSTAR}} band with an optical depth greater than 0.15. An iron emission line is detected in both the {\it{NuSTAR}} spectra with an equivalent width of about 125 eV. We looked at the dependence of the spin-up rate on the luminosity and estimated the magnetic field strength from that, which came out to be much higher compared to other known BeXRB pulsars. Lastly, we discussed the inadequacy of the torque-luminosity relation for determination of magnetic field strength of neutron stars.
\end{abstract}




\section{Introduction}
In X-ray binary pulsars (XBPs) the neutron star accretes matter from its companion and the in-falling matter after crossing the Alfven radius (radius inside which the magnetic stress influences the flow in the accretion disk) is channeled to the magnetic field lines and finally reaches the poles of the neutron star. The gravitational potential energy of the accreted matter powers X-ray radiation and the X-ray luminosity is governed by the mass accretion rate. The angular momentum of the accreted matter is added to the angular momentum of the neutron star, causing it to rotate faster. Therefore the spin-up rate and X-ray luminosity are supposed to be correlated, and this relation depends on various factors like mass, radius, magnetic field of the neutron star, mode of accretion, etc. According to the standard model of accretion, $\dot{\nu} \propto L_{X}^{6/7}$ \citep{1979_Ghosh}, where  $\dot{\nu}$ is the spin frequency change rate (=$-\frac{\dot{P}}{P^2}$, $P$ is the spin period).
\par
In Be X-ray binaries the neutron star accretes matter from its companion's slow, dense equatorial stellar outflow and an accretion disk is often present during giant outbursts. During these giant outbursts, the X-ray source shows large changes in the X-ray luminosity. The rapid spin up episodes during such outbursts make Be X-ray binaries excellent candidates to establish a quantitative comparison of theoretical predictions with minute observations. The luminosity and the spin frequency derivatives during large outbursts with $L_{X} \geq 1 \times 10^{37} erg s^{-1}$ were found to be related as $\dot{\nu} \propto L_X^{\alpha}$ $(\alpha \simeq 1)$ \citep{1989_Parmar, 1997...Bildten, 2017_Sugizaki}. This proportionality relation validates the expected theoretical XBP spin up models of magnetospheric interaction with accretion disk. The giant outbursts also provide an excellent opportunity to characterize the broadband pulsation and spectral properties, especially with observatories like {\it{NuSTAR}}. Detection of a Cyclotron Resonance Scattering Feature (CRSF) in the X-ray spectrum helps to determine the magnetic field strength of the neutron star unambiguously. In this paper, we study one such Be XBP called GRO J2058+42.        
\par
The transient 198 second X-ray pulsar GRO J2058+42 (a.k.a. CXOU J205847.5+414637) was discovered during a giant outburst with BATSE onboard the Compton Gamma Ray Observatory on September 14, 1995 \citep{1995_Wilson,1998_Wilson}. The outburst lasted for about 46 days until October 30, 1995, and spun up to a period of 196 seconds with the pulsed flux peaking at 140 mCrab (20-50 keV) on September 27, 1995. The initial giant outburst was followed by a series of normal outbursts with pulsed flux peaking at 15-20 mCrab (20-50 keV) occurring at $\sim$ 110 days intervals. BATSE also detected shorter and weaker outbursts with peak-pulsed fluxes of $\sim$ 8 mCrab in between the 15-20 mCrab outbursts. These normal and weaker outbursts were similar in intensity when observed with \textit{RXTE}-ASM indicating an orbital cycle of 55 days \citep{1997_Corbet}. An alternate explanation was that the normal and weaker outbursts are related to two different accretion mechanism during the periastron and apastron passage and the orbital cycle is of $\sim$110 days \citep{1998_Wilson}. The occurrence of periodic outbursts after a giant outburst indicates the companion of GRO J2058+42 is likely to be a Be star.
\begin{figure}
    \centering
    \includegraphics[scale=0.35, trim={0 1.8cm 0 0}, angle=-90]{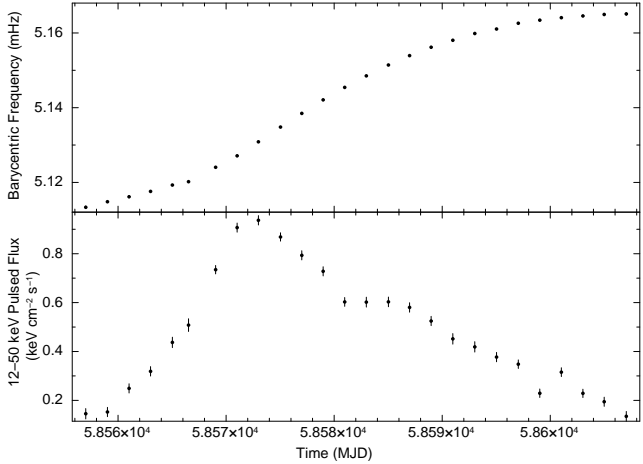}
    \caption{Top: barycentric spin-frequency history measured with GBM.The typical error associated with each pulse frequency measurement is of the order of 10$^{-4}$ mHz, which is too small to be visible in this plot. Bottom: pulsed flux measured with GBM in 12-50 keV energy band.}
    \label{fig:spin_flux}
\end{figure}
With the optical and X-ray observation carried out during quiescence, GRO J2058+42 was proven to be a Be X-ray binary \citep{2005_Reig,2005_Wilson}. \cite{2005_Reig} identified the companion as an O9.5-B0 IV-Ve star with V $\simeq$ 14.9 mag and R $\simeq$ 14.2 mag at a distance of $\sim$ 9 $\pm$ 1 kpc, which is in agreement with the distance range of 7-16 kpc estimated from accretion torque during the first giant outburst \citep{1998_Wilson}. On May 2, 2008 another outburst activity from GRO J2058+42 was detected with {\it{Swift}}-BAT hard X-ray transient monitor in 15-50 keV energy band. The BAT flux was similar to the normal outbursts and it was not accompanied by any giant outburst. 
\par
 Recently, a second giant outburst from the source has been observed with {\it{Swift}}-BAT and {\it{Fermi}}-GBM in the mid March of 2019. {\it{Swift}}-BAT triggered at this source on March 22 \citep{2019_Barthelmy} and on March 27 \citep{2019_Kennea,2019_Lien} reporting a renewed activity. \cite{2019_Malacaria} reported the GBM detection of the source, going through a bright outburst starting on March 14, 2019 reaching to a peak luminosity similar to the previous giant outburst in 1995 observed by BATSE. It was observed by GBM for $\sim$ 52 days. In this period, {\it{NuSTAR}} carried out two target of opportunity (TOO) observation on March 25 and April 11. 
\begin{figure}
    \centering
    \includegraphics[width=\columnwidth , scale=1.2, angle=-90]{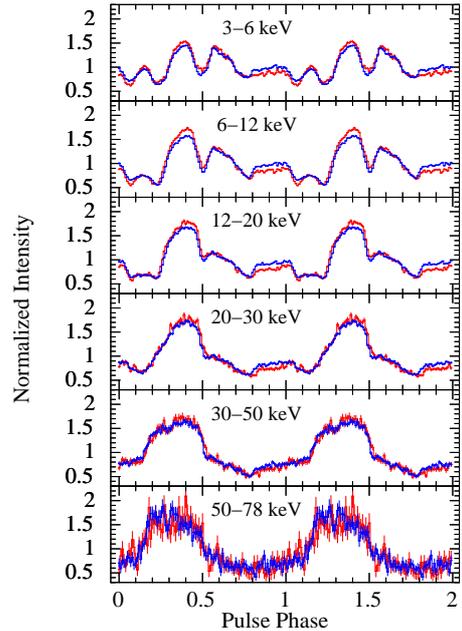}
    \caption{Energy resolved pulsed profiles from {\it{NuSTAR}} observations with Obs ID 90501313002 (red) and Obs ID 90501313004 (blue) folded with $P_{spin}=$ 195.25 seconds and 194.14 seconds respectively. }
    \label{fig:pulse profile}
\end{figure}

Among the bright transient Be X-ray binaries, GRO J2058+42 is a relatively poorly studied source. Using the GBM and {\it{NuSTAR}} observation of the recent second giant outburst of this source, we have looked at its broadband X-ray spectrum and searched for the CRSF feature, if any, and studied the spin-up rate and pulsed X-ray luminosity to estimate the magnetic field (B field) strength of the pulsar. The spin-up characteristics was previously studied by \cite{1998_Wilson} from its first giant outburst observation but the distance of the binary was still unknown. So it was not possible to estimate the B field earlier. We have also analyzed the pulse profiles in broad energy band from two {\it{NuSTAR}} observations to look at their pulse shapes in order to find any change in the accretion mode.
\begin{figure*}
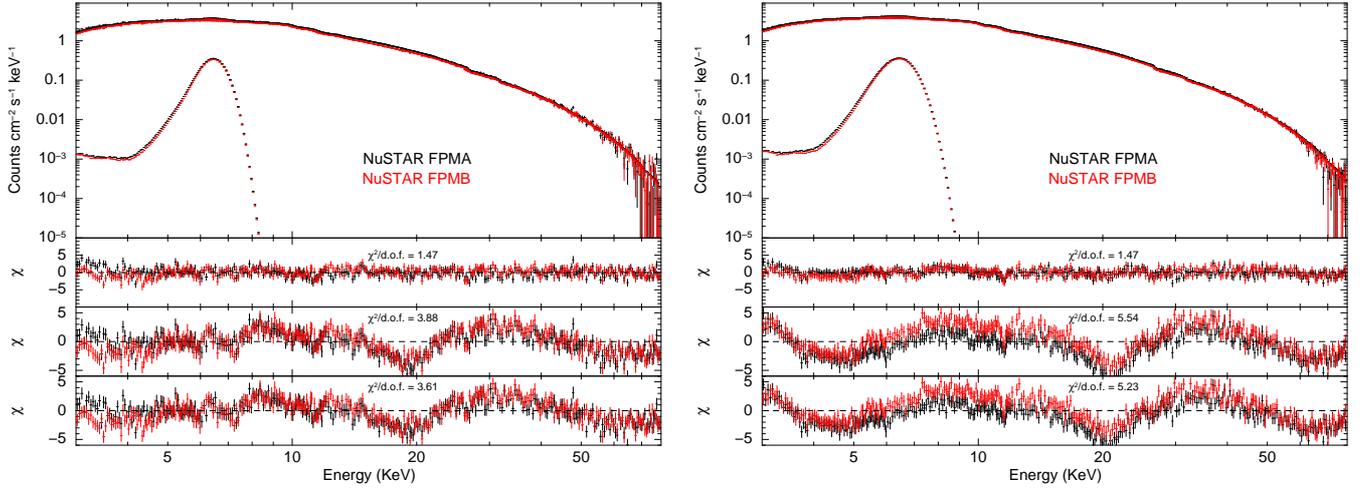

	\includegraphics[scale=0.37, trim={0 4.0cm 0 0.5cm},  angle=-90]{002.eps}
	\includegraphics[scale=0.37, trim={0 0 0 1.8cm},  angle=-90]{004.eps}
    \caption{The first panel are the overlaid X-ray spectra of FPMA and FPMB from NuSTAR observations with ObsID 90501313002 (left) and ObsID 90501313004 (right) modelled with absorbed power law and FDCUT along with an emission line. The second, third and fourth panel represent the (data-model)/error (i.e. $\chi$) when the cut-off energy component is FDCUT, HIGHECUT and NEWHCUT respectively. These fits do not include a gaussian absorption line in the model. }
    \label{fig:spectra}
\end{figure*}
\section{Observations} 
\subsection{\textbf{{\textit{NuSTAR}}} observations}
The Nuclear Spectroscopic Telescope Array ({\it{NuSTAR}}) is first focusing hard X-ray mission that detects X-rays in 3 to 78 keV energy band with two co-aligned telescopes with focal plane modules FPMA and FPMB \citep{2013_Harrison}. The telescopes have 18" FWHM imaging resolution with a characteristic spectral resolution of 400 eV FWHM at 10 keV and a temporal resolution of 2 $\mu$s. TOO observation of GRO J2058+42 were made on Mar 25, 2019 first time for 20.4 ks (ObsID 90501313002, $T_{start}=$58567.30 MJD) and then again on April 11, 2019 for 38.6 ks (ObsID 90501313004, $T_{start}=$58584.01 MJD).  
\par
To process and filter the preliminary {\it{NuSTAR}} data we have used HEASOFT version 6.25 in which NuSTARDAS pipeline version 1.8.0 is installed as a sub-package. First using standard NuSTARDAS pipeline, we extracted the clean event files. Then using DS9 version 8.0.1, 120" radius region centering the source was selected. It was used to extract the source photons in order to create the source lightcurve and spectra. Another circular region of 120" away from the source in the FoV was selected to create background lightcurve and spectra. The response matrix files and ancillary files were generated using CALDB version 1.0.2. 

\subsection{\textbf{{\textit{Fermi}}}-GBM observations}
The Fermi Gamma-ray Space Telescope has two main instruments on-board, the Large Area Telescope (LAT) and the Gamma-ray Burst Monitor (GBM). GBM is composed of 14 detectors: 12 Sodium Iodide (NaI) detectors and two Bismuth Germanate (BGO) detectors \citep{2009_Meegan} and is sensitive within the energy range between 8 keV to 40 MeV.
In this analysis we have used the pulse frequency and 12-50 keV pulsed flux measurements with the
{\it{Fermi}}-GBM \citep{2009_Finger, Camero_Arranz_2009}. The measurement technique is discussed thoroughly in \cite{2020_Malacaria}.
\par
GRO J2058+42 was detected with  GBM from Mar 14, 2019 to May 4, 2019 for a period of about 50 days during its most recent outburst.
Figure \ref{fig:spin_flux} represents the barycentric pulse frequency and pulsed flux profile with time \footnote{\url{https://gammaray.msfc.nasa.gov/gbm/science/pulsars/lightcurves/groj2058.html}}. The error bars in figure \ref{fig:spin_flux} stands for the associated uncertainties in these measurements due to statistical limitations and time intervals. The errors in pulse frequency measurements are about four orders of magnitude smaller than the pulse frequency values and are smaller than the symbols used in the figure \ref{fig:spin_flux}.

\section{Data Analysis}
\subsection{Timing Analysis} 
We created source and background lightcurves from the two {\it{NuSTAR}} observations with a bin time of 1 second for both FPMA and FPMB. For each observation, we added the lightcurves from FPMA and FPMB after background subtraction.
We determined the pulse period ($P_{spin}$) during the two observations by the method of epoch folding and chi-square maximization using \textbf{efsearch} v1.1 (XRONOS v5.22). The pulse periods at reference epoch MJD 58567.30 and MJD 58584.01 were found to be 195.25 $\pm$ 0.02 seconds and 194.14 $\pm$ 0.02 seconds respectively. These two periods are consistent with the pulse period evolution recorded with {\it{Fermi}}-GBM during the outburst.
\par
Then, following the same method we generated 6 background-subtracted lightcurves in 3-6 keV, 6-12 keV, 12-20 keV, 20-30 keV, 30-50 keV and 50-78 keV energy band for each of the two observations and folded them with their respective pulse periods. Then we aligned and overlaid these energy resolved pulse profiles from two observation on top of each other (Figure \ref{fig:pulse profile}) to observe any change in the pulse shapes. They do not look significantly different from each other.
\par
The pulse shapes has strong energy dependence. At lower energies the pulse profile has at least four peaks and the two main peaks merge at higher energies. The energy dependence of the pulse shapes are consistent with the previous \textit{RXTE}-PCA observation in November 28, 1996 \citep{1998_Wilson}.
\subsection{Spectral Analysis}
Source and background pulse phase-averaged spectra and response files were created from the two {\it{NuSTAR}} observations using \textbf{nuproducts} command in the 3-78 keV energy band. We have fitted FPMA and FPMB spectra with the same spectral model but allowed the relative normalization of the detectors to vary.

\begin{table*}
\caption{ Spectral parameters with quoted errors for 90\% confidence limits. }
\label{spectral_parameter}
  \centering
 \begin{tabular}{|c|c|c|c|c|c|c|c|}
 \hline
 &&&&&&\\
 
 parameter && ObsID 90501313002 &&& ObsID 90501313004 &\\
 \hline
       & FD-CUT & HIGHECUT & NewHcut  & FD-CUT & HIGHECUT   & NewHcut      \\
       &&&&&&\\
        \hline
          $N_H^a$ (fixed)      & 6.19 $\times$10$^{-1}$ & 6.19 $\times$10$^{-1}$ & 6.19 $\times$10$^{-1}$ & 6.19 $\times$10$^{-1}$ & 6.19 $\times$10$^{-1}$ & 6.19 $\times$10$^{-1}$ \\
     &&&&&&\\
     Flux$^b$  & 3.67$_{-0.01}^{+0.01}$ & 3.68$_{-0.01}^{+0.01}$ & 3.68$_{-0.01}^{+0.02}$ & 4.42$_{-0.01}^{+0.01}$ & 4.43$_{-0.01}^{+0.01}$ & 4.42$_{-0.01}^{+0.01}$ \\
  
 \hline
     &&& continuum &&&&\\
 \hline
     Photon Index ($\Gamma$)         & 0.88$_{-0.01}^{+0.01}$ & 0.95$_{-0.01}^{+0.01}$ & 0.95$_{-0.01}^{+0.01}$ &0.83$_{-0.01}^{+0.01}$ & 0.89$_{-0.01}^{+0.01}$ & 0.86$_{-0.01}^{+0.01}$ \\ 
     &&&&&&\\
     $\Gamma_{norm}^c$               &6.27$_{-0.05}^{+0.05}$ & 6.11$_{-0.08}^{+0.08}$ & 6.11$_{-0.09}^{+0.08}$ & 7.04$_{-0.05}^{+0.05}$ &6.47$_{-0.09}^{+0.08}$ & 6.40$_{-0.09}^{+0.09}$ \\
     &&&&&&\\
     
     $E_{cut}$ (keV) & 27.72$_{-0.60}^{+0.58}$ &24.31$_{-0.36}^{+0.34}$ &24.99$_{-0.79}^{+1.14}$ &24.79$_{-0.60}^{+0.58}$ &24.26$_{-0.34}^{+0.35}$ &24.47$_{-0.50}^{+0.52}$\\
     &&&&&&\\
     $E_{fold}$ (keV)  & 11.76$_{-0.20}^{+0.20}$ &14.83$_{-0.27}^{+0.28}$ & 14.33$_{-0.64}^{+0.49}$ &12.71$_{-0.16}^{+0.16}$ &14.47$_{-0.23}^{+0.23}$ & 14.17$_{-0.28}^{+0.28}$ \\
     &&&&&&\\
     
     \hline
      &&& emission lines &&&&\\
     \hline
       Fe-K$\alpha$ line energy (keV) &6.46$_{-0.03}^{+0.03}$ & 6.46$_{-0.03}^{+0.03}$ & 6.46$_{-0.03}^{+0.03}$ & 6.45$_{-0.04}^{+0.04}$ & 6.43$_{-0.04}^{+0.04}$ & 6.43$_{-0.04}^{+0.04}$\\
       &&&&&&\\
       $\sigma_{K\alpha}$ (keV) & 0.36$_{-0.04}^{+0.04}$ & 0.42$_{-0.05}^{+0.05}$ & 0.42$_{-0.04}^{+0.05}$ & 0.48$_{-0.05}^{+0.06}$ & 0.54$_{-0.06}^{+0.06}$ & 0.52$_{-0.06}^{+0.06}$ \\
       &&&&&&\\
       Equivalent width (eV)  & 114$_{-10}^{+8}$ & 136$_{-8}^{+9}$ & 133$_{-9}^{+9}$ & 129$_{-11}^{+10}$ & 146$_{-10}^{+10}$ & 140$_{-9}^{+9}$ \\
       &&&&&&\\
       
       $E_{gabs}$  (keV)    & -- & 24.41$_{-0.43}^{+0.43}$ & 25.12$_{-0.89}^{+1.25}$ & -- &  24.48$_{-0.39}^{+0.40}$ & 24.71$_{-0.55}^{+0.57}$ \\
       &&&&&&\\
       $\sigma_{gabs}$ (keV) & -- & 5.62$_{-0.30}^{+0.32}$ & 6.14$_{-0.40}^{+0.48}$ & -- & 6.50$_{-0.39}^{+0.40}$ & 6.95$_{-0.30}^{+0.31}$ \\
        &&&&&&\\
        $strenth_{gabs}$ & -- & 3.67$_{-0.34}^{+0.38}$ & 4.23$_{-0.58}^{+0.90}$  & -- & 5.25$_{-0.42}^{+0.46}$ & 5.83$_{-0.53}^{+0.58}$ \\
        &&&&&&\\
\hline
      &&& statistic &&&&\\
  \hline
     
     $\chi^{2}/d.o.f.$   &1.47$^d$ & 1.58$^e$ (3.88$^d$) & 1.56$^e$ (3.61$^d$) & 1.13$^d$  & 1.19$^e$ (5.54$^d$)  & 1.14$^e$ (5.23$^d$)\\
    
&&&&&&\\    
\hline
      &&& relative normalization &&&&\\
\hline
    FPMB  & 1.023$_{-0.003}^{+0.003}$ & 1.023$_{-0.003}^{+0.003}$ & 1.023$_{-0.003}^{+0.003}$ & 1.028$_{-0.002}^{+0.002}$ & 1.028$_{-0.002}^{+0.002}$ & 1.028$_{-0.002}^{+0.002}$ \\
     &&&&&&\\

 \hline
 
  \end{tabular}
 \begin{flushleft}
 $^a$ : XSPEC normalization , units of 10$^{22}$ atoms cm$^{-2}$\\
 $^b$ : XSPEC normalization , units of 10$^{-9}$ ergs cm$^{-2}$ sec$^{-1}$ \& in the 3-78 keV range\\
 $^c$ : XSPEC normalization , units of 10$^{-2}$ photons cm$^{-2}$ s$^{-1}$ at 1 keV\\
 $^d$ : without gabs\\
 $^e$ : with gabs
 
  \end{flushleft}
 \end{table*}

First we tried various continuum models used for HMXB pulsars like cutoff power-law, high energy cutoff power-law \citep[HIGHECUT;][]{1983_White}, NewHcut \citep[a third order polynomial function with continuous derivatives;][]{2000_Burderi}, Fermi Dirac cutoff power-law \citep[FD-CUT;][]{1986_Tanaka}, Thermal Comptonization model \citep[CompTT;][]{1994_Titarchuk} to fit the spectra. The HIGHECUT, NewHcut and  Fermi Dirac cutoff power-law spectral models described the source spectrum well with physically acceptable parameters. For all these three cases, Fe-K$\alpha$ line at 6.4 keV is present in both the spectra with an equivalent width of about 125 eV. However, when fitted with the HIGHECUT and NewHcut cut-off power law model, the spectra required an additional absorption component (gaussian absorption line \footnote{\url{https://heasarc.gsfc.nasa.gov/xanadu/xspec/manual/node240.html}}) which indicated possible presence of a CRSF feature at around 25 keV. But in all such cases, the cyclotron energy is almost same as the cut-off energy, which implicate the CRSF feature to be an artifact. FD-CUT model does not show presence of any negative residual. and provided the best fit to the spectra with minimum $\chi^2$. While fitting the spectra, we considered a component {\it{phabs}} to account for the absorption in the interstellar medium along our line of site. But the value of $N_H$ came out to be smaller than the galactic $N_H$ and it could not be constrained. Therefore, we have included neutral hydrogen absorption fixed to the Galactic value of $N_H$= 6.19 $\times$ 10$^{21}$ cm$^{-2}$ calculated using HEASARC tools \footnote{\url{https://heasarc.gsfc.nasa.gov/cgi-bin/Tools/w3nh/w3nh.pl}}. The spectral parameters for these models are tabulated in Table \ref{spectral_parameter}.
\newline
Figure \ref{fig:spectra} shows the overlaid FPMA and FPMB spectra from two observations by {\it{NuSTAR}}. They are fitted with {\it{constant*phabs(powerlaw*fdcut + gaussian)}} where the {\it{constant}} is to allow for a relative normalization for FPMA and FPMB. It is kept as 1 for FPMA and varied for FPMB. {\it{phabs}} $N_H$ is to account for the galactic value. A {\it{gaussian}} component is used to fit the Iron line at 6.4 keV and {\it{powerlaw*fdcut}} is used to fit the continuum. The second, third and fourth panel in each side of Figure \ref{fig:spectra} are the $\chi$ when the cutoff energy components are FDCUT, HIGHECUT and NEWHCUT respectively.
\par
 We also fitted the broadband spectra with the best fit model and included an extra absorption component ({\it{gabs}}) at a particular energy and repeated this for every energy in the {\it{NuSTAR}} band. We determined the upper limit on the optical depth each time and presented it in Figure \ref{fig:optical depth}. To determine the upper limit of any absorption line in the spectrum, the width of the line was taken to be 1 keV and 2 keV in the energy bands of 4-10 keV and 11-78 keV respectively. Since the spectrum is only up to 80 keV and statistics is poor at higher energy, it is not well constrained beyond 60 keV.

\section{Accretion torque and estimation of the magnetic field strength}
\subsection{Spin change rate and luminosity measurement}
The pulsed frequencies ($\nu_{s}$) measured with {\it{Fermi}}-GBM clearly show a rapid spin-up of the pulsar during the second giant outburst (Figure \ref{fig:spin_flux}). The total increase in frequency during the second  outburst is about 0.05 mHz which is almost similar to 0.06 mHz during the first giant outburst. In total there were 26 $\nu_{s}$ measurements taken roughly in every 2 day intervals. We took first three consecutive $\nu_{s}$ and fitted them w.r.t. time with a linear function. While fitting, we took into account the uncertainties associated with the pulse frequency measurements and because the integration time for each pulsation value is 2 days, the uncertainties associated with time is taken to be 1 day. We achieved the best fit using $\chi^2$ minimization technique and determined the spin up rate from the slope of the linear function for 6 days interval, as each pulse frequency measurement is taken in every two days interval. We estimated the 1$\sigma$ statistical error on spin-up rate ($\dot{\nu}_{s}$). We repeated this process for the next three $\nu_{s}$ and so on. Therefore, we had 8 spin-up rates ($\dot{\nu}_{s}$) from 24 spin frequencies ($\nu_{s}$) and we did not use the last 2 $\nu_{s}$.



\par
From the {\it{NuSTAR}} data we estimated total flux (3-78 keV) at the time of the {\it{NuSTAR}} observations. 
We determined a conversion factor by which a pulsed flux measurement is related to the total flux from the {\it{NuSTAR}} observation at a particular time. Then we multiplied all the 26 pulsed flux detected with {\it{Fermi}}-GBM with this conversion factor to convert them into the total flux. To make a simultaneous sampling of the luminosity and spin-up rate measurements, we have taken the average value of the total flux over the same 6 days intervals which are used to determine a particular $\dot{\nu}_{s}$. We then had 8 spin-up rates and total flux measurements during the outburst. We multiplied the uncertainties on the pulsed flux values by the same conversion factor and estimated the propagated error on the average value of the total fluxes. Then using the distance estimation of 9 kpc \citep{2005_Reig}, we have calculated the luminosity, along with its uncertainty. A plot of the spin-up rate against bolometric luminosity is shown in Figure \ref{fig:nudot_luminosity}. We have made a power law fitting of $\dot{\nu}_{s}$ and luminosity to check the relation between them. 

\begin{figure}
    \centering
    \includegraphics[scale=0.325, trim={0 1.8cm 0 0}, angle=-90]{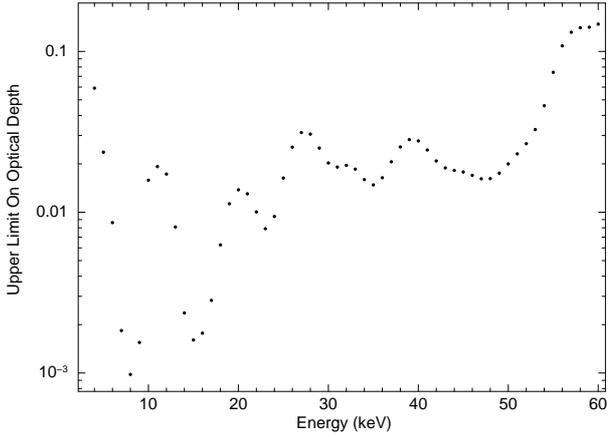}
    \caption{The upper limit on optical depth of a gaussian absorption line as a function of energy. This figure was generated from the \textit{NuSTAR} observation with the ObsID 90501313002.  }
    \label{fig:optical depth}
\end{figure}

\subsection{$\dot{\nu}_{s}$ and $L$ relationship}
The spin-up rate ($\dot{\nu}_{s}$) and the luminosity $L$ are known to be correlated in transient X-ray pulsars \citep{2017_Sugizaki}. The disk-magnetosphere interaction model of a rotating neutron star which is spinning up by accreting mass from it's companion via a Keplerian disk, with a moment of inertia $I$, holds the following relation approximated to the non-relativistic limit \citep{1979_Ghosh, 2017_Sugizaki}.\\
\begin{equation}
\dot{\nu}_{12} = 2.0\,
  n\zeta^{1/2}
  \mu_{30}^{2/7} R_{6}^{6/7} M_{1.4}^{-3/7}I_{45}^{-1} 
  L_{37}^{6/7} 
\label{equ:nudot2}
\end{equation}
where $\dot{\nu}_{12}$, $\mu_{30}$, $R_{6}$, $M_{1.4}$, $I_{45}$ are the spin frequency derivative, magnetic dipole moment, radius, mass and the moment of inertia of the neutron star given in the units of $10^{-12}$ Hz s$^{-1}$, $10^{30}$ G cm$^3$, $10^6$ cm, 1.4 $M_{\odot}, 10^{45}$ g cm$^2$ respectively. Also, the X-ray luminosity is termed as $L_{37}$ in the unit of $10^{37}$ erg sec$^{-1}$.
$n$ and $\zeta$ are both dimensionless parameters. $n$ is a function of "fastness parameter"(i.e. the the ratio of the neutron star's angular frequency to that of the disk at the radius $r_{0}$ at which the magnetic barrier terminates it). $\zeta$ is the factor by which $r_{0}$ differs from the Alfven radius $r_a$ (i.e. $r_{0}=\zeta r_a$). According to \cite{1979a_Ghosh, 1979_Ghosh} model, under slow-rotator condition $n$ $\sim 1.39$ and $\zeta \sim 0.52$. Therefore, equation~(\ref{equ:nudot2}) reduces to \citep{2017_Sugizaki},\\
\begin{figure}
    \centering
    \includegraphics[width=\columnwidth ,trim={0 1.8cm 0 0}, angle=-90, scale=0.75]{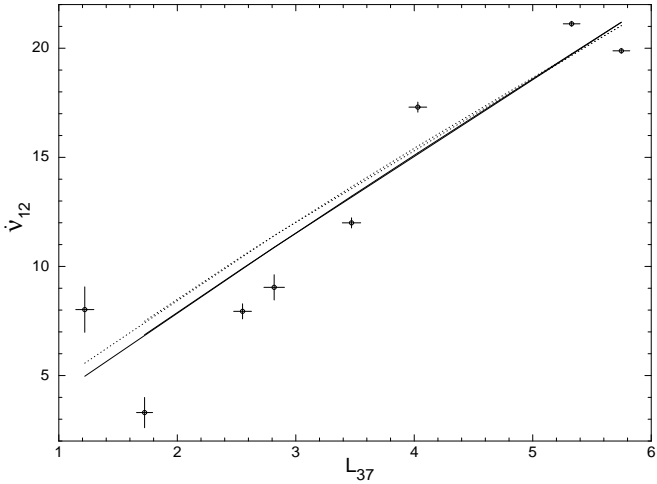}
    \caption{Plot of spin change rate and luminosity. The dotted line has power-law index of 6/7 and the solid line represents the best power-law fit of the data points which gives a power-law index of 0.93.}
    \label{fig:nudot_luminosity}
\end{figure}

\begin{equation}
\dot{\nu}_{12} = k L_{37}^{\alpha}  
\label{equ:nudot3}
\end{equation}
where $k = 2.0\, \mu_{30}^{2/7}$ and $\alpha = 6/7 \simeq 0.857$.\\

For nominal values of $R_{6}$ = $M_{1.4}$ = $I_{45}$ = 1, measurements of the $\dot{\nu}_{12}$ vs. $L_{37}$ provides as a rough estimation of the magnetic dipole moment of the pulsar. Observationally, $\alpha$ ranges between 0.85-1. In Figure \ref{fig:nudot_luminosity}, the solid line is the best fit with $\alpha = 0.93$ and the dotted line is the fit for the theoretical value of $\alpha$ (i.e. 0.857). These two values are almost equal. We estimated the value of $k = 4.13 \pm 0.25$ and $\alpha = 0.93 \pm 0.04$ within 90$\%$ confidence range. From the best fit the $\dot{\nu}_{12}$ and $L_{37}$ relation turns out to be  \\
\begin{equation}
\dot{\nu}_{12} = 4.13\, L_{37}^{0.93} \\
\label{equ:nudot4}
\end{equation}

As, $\alpha$ is close to the theoretical value, we take\\
\begin{equation}
\Rightarrow 2.0\, \mu_{30}^{2/7} = 4.13\\
\Rightarrow \mu_{30} = 12.6
\label{equ:nudot5}
\end{equation}

\begin{figure}
    \centering
    \includegraphics[scale=0.35, angle=-90]{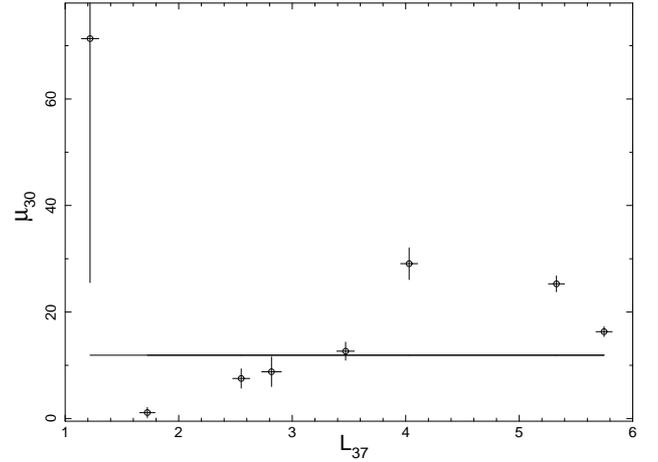}
    \caption{Plot of $\mu_{30}$ vs. $L_{37}$. The solid line is best fit with a constant which gives $\mu_{30}$ = 11.89.}
    \label{fig:mu_L}
\end{figure}

We also calculated the value of $\mu_{30}$ for each set of $\dot{\nu}_{12}$ \&  $L_{37}$ and also the error on it from the propagated error of $\dot{\nu}_{12}$ \&  $L_{37}$. Figure \ref{fig:mu_L} shows the variation in $\mu_{30}$ with luminosity. We modelled it with a constant for all the values of $\mu_{30}$.
The value of the constant came out to be 11.89 which is shown by the solid line in Figure \ref{fig:mu_L}.

\subsection{Estimation of the surface magnetic field}
The magnetic moment $\mu_{30}$ can be expressed in terms of surface magnetic field strength $B_{12}$ and the radius of the pulsar $R_6$ as,\\
\begin{equation}
   \mu_{30} = \frac{1}{2} B_{12} R_{6}^{3}\phi(x) 
   \label{equ:nudot6}
\end{equation}
where $\phi (x)$ is a correction factor \citep{1983_Wasserman} on surface redshift parameter, x ($= \frac{Rc^2}{2GM}$) which is the ratio of the pulsar radius to the Schwarzschild  radius. For a typical neutron star this ratio is $\sim$ 2.4 and the correction factor $\phi (x) \sim 0.68$. So equation \ref{equ:nudot6} becomes,
\begin{equation}
    B_{12} = \frac{2\times \mu_{30}}{0.68} 
\end{equation}
For $\mu_{30} \simeq 12.6$ , $B_{12} \simeq 37$. So, the surface magnetic field strength estimated from the spin-up characteristics of GRO J2058+42 during its second outburst is $\simeq 37 \times 10^{12}$ Gauss. It is rather large compared to the surface magnetic field estimated from the cyclotron line in most accreting X-ray pulsars in Be X-ray binaries.

\begin{table}
\caption{ Comparison of estimated strength of surface magnetic field of nine transient pulsars from the $k^*$ and from observed cyclotron energy. }
\label{B12}
  \centering
 \begin{tabular}{|c|c|c|}
 \hline
 &&\\
  
  Source & $B_{12}$ from $k^*$ & $B_{12}$ from observed $E_{cyc}$\\
  \hline
  
 4U 0115+63   &	1.15$\times 10^{-2}$	&	1.81\\
  X 0331+53   &	3.76$\times 10^{-5}$	&	3.50\\
 RX J0520.5-6932   &	6.07$\times 10^{-2}$	&	3.56\\
 H 1553-542   &	5.21$\times 10^{-1}$	&	3.08\\
  XTE J1946+274  &	1.23                	&	3.95\\
 KS 1947+300   &	5.33                	&	1.38\\
 GRO J1008-57   &	1.41$\times 10^{-1}$	&	8.58\\
  A 0535+262  &	8.03$\times 10^{-1}$	&	5.28\\
  GX 304-1  &	2.51$\times 10^{-1}$	&	6.06\\
 
 &&\\
 \hline
 \end{tabular}
 \begin{flushleft}
 $^*$ : Equation \ref{equ:nudot3}
 \end{flushleft}
 \end{table}

\begin{figure}
    \centering
    \includegraphics[scale=0.35, trim={0 1.8cm 0 0}, angle=-90]{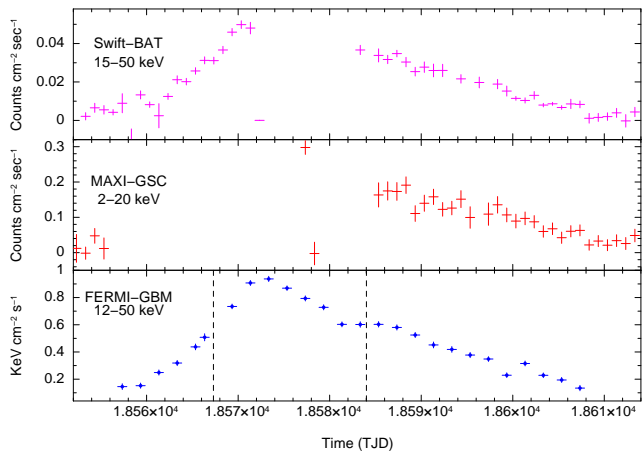}
    \caption{The lightcurve of GRO J2058+42 during it's second outburst observed by {\it{Swift}}-BAT in the top panel and by {\it{MAXI}}-GSC in the middle panel. The bottom panel shows the pulsed fluxes observed by {\it{Fermi}}-GBM and the two black vertical lines represent the start time of the two {\it{NuSTAR}} observation. }
    \label{fig:GBM_MAXI_BAT}
\end{figure}

\section{Discussions}
  Be X-ray binary systems usually have a wide eccentric orbit and a gaseous circumstellar geometrically thin disk in Keplerian motion around the equator of the Be star. Such systems are often dormant or show weak outbursts in every orbit. However, sometimes the outflow from the companion star is enhanced and when the neutron star crosses the disk, large outbursts for weeks may occur. The transient Be X-ray binary GRO J2058+42 has gone through its second known giant outburst in late March of 2019. The pulse frequency and pulsed flux were measured with {\it{Fermi}}-GBM throughout the outburst and there were two Target of Opportunity observations with the {\it{NuSTAR}} during the outburst. We have carried out timing and spectral analysis of the {\it{NuSTAR}} data along with a study of its spin frequency evolution with the luminosity and estimated the strength of the magnetic field of the pulsar. 
  \par
From the first \textit{CGRO}-BATSE observation of the giant outburst \citep{1998_Wilson}, pulse profiles of GRO J2058+42 were studied during the fast rise and then slower rise followed by the decay of the outburst and significant evolution in the pulse profile was witnessed from a fast rise to the decay. The pulse profile is representative of the beaming pattern. Hence it is a tool to observe any change in the accretion mode. The pulse profiles from the first outburst showed clockwise transformation of pulses. However, our phase aligned pulse profiles from the two {\it{NuSTAR}} observations, one during the rise and the other during the fall of the recent outburst look quite similar to each other (Figure \ref{fig:pulse profile}). On the other hand, in terms of energy dependence, energy resolved pulse profiles from the previous and the recent outburst are in agreement. In the soft X-ray band there are at least four different pulse components at low energy which evolve to a single peaked profile at high energy (> 30 keV). Multiple peaks in low energy and single peaked pulses in higher X-ray energy band has also been seen in other pulsars like GRO J1008-57 \citep{2011_Naik} and GX 304-1 \cite{2011_Devasia}. 
\par
Previously, \cite{1998_Wilson} studied the relationship between its spin-up rate and X-ray flux from its first outburst observation. The value of $\alpha$ (equation \ref{equ:nudot3}) was found to be $\sim 1.2$, which is slightly higher than that is reported here for the second outburst.
The X-ray luminosity shown in Figure \ref{fig:nudot_luminosity} is calculated from the hard X-ray pulsed emission assuming that the X-ray spectral shape and pulsed fraction remained unchanged through out the period of Fermi observation. The medium energy X-ray light curve obtained with \textit{MAXI}-GSC (2-20 keV), the hard X-ray light curve obtained with \textit{Swift}-BAT (15-50 keV) and the pulsed flux of the source measured with \textit{Fermi}-GBM during the outburst are shown in Figure \ref{fig:GBM_MAXI_BAT}. We have carried out separate analysis comparing the three light curves. We checked the hardness ratio using ratio of the count rates with time from 
\textit{Swift}-BAT (15-50 keV) and \textit{MAXI}-GSC (2-20 keV). We did not observe any significant
change in the spectral shape. Also, we looked at the ratio of the count rate from the
\textit{Swift}-BAT and the pulsed flux from the \textit{Fermi}-GBM observations during the entire
outburst. The ratio was almost equal during the entire period, which indicated that there
was no significant change in the pulsed fraction during this period.
Within error bars the X-ray spectral shape and hard X-ray pulse fraction remained the same during the outburst. Another unknown factor that may contribute to the pulse period and period derivative measurements is the orbital motion of the pulsar. As the orbital parameters of GRO J2058+42 are unknown, it is not possible to correct the data for orbital motion.
\par
From the $\dot{\nu}_{12}$ vs. $L_{37}$ relationship, we calculated the value of the magnetic moment of the neutron star and from that we estimated the magnetic field strength which came out to be $\simeq 37 \times 10^{12}$ Gauss. The high magnetic field in neutron stars causes the formation of the CRSF in the X-ray spectra. This broad absorption like feature in the energy spectrum, arises due to the X-ray photons in the accretion column scattered by the plasma electron quantized in the Landau levels. The magnetic field strength and the fundamental cyclotron energy is related as $B_{12}= \frac{1}{\sqrt{1-x^{-1}}}\frac{E_{cyc}}{11.6keV}$. For Be XRPs the CRSF is observed within a energy range of 12--76 keV \citep{2017_Sugizaki}. The energy spectra of GRO J2058+42 does not show any CRSF in 3-78 keV energy band. From the estimated $B_{12} \simeq 37$, we get $E_{cyc} \simeq 328$  keV. So, the cyclotron line is expected to be outside the energy band of the X-ray spectrum measured with {\it{NuSTAR}}.
\par
The $B_{12}$ and $E_{cyc}$ estimated for GRO J2058+42 from the accretion torque is much larger than that of other Be XRPs. One of the reason can be an under estimation of the total X-ray luminosity. While estimating the total flux from the pulsed flux, we considered the energy range of {\it{NuSTAR}} i.e. 3-78 keV. So, the bolometric X-ray luminosity in 0.1-100 keV band should be higher. 
With the best fit {\it{NuSTAR}} model, for both the observations, we calculated the 0.1-100 keV flux, which is larger than the 3-78 keV flux by only 3-4\%. However, the surface magnetic field is high enough that we do not see the CRSF in the {\it{NuSTAR}} band, which indicates to a $B_{12}\geq 9$ and it is higher than the magnetic fields confirmed for HMXBs \citep{2012_Cab}. There are few other Be XRPs like EXO 2030+375 \citep{2013_Naik}, 2S 1417-624 \citep{2018_Gupta} and GS 0834-430 \citep{2013_Miya}, for which no CRSF feature has been found. So, it can be a possibility that some of the Be XBPs may have a high magnetic field of the order of $10^{13}$ Gauss. 
Despite a  non-detection of any cyclotron line in the spectra, we are not entirely rejecting the idea of presence of it and provided an upper limit of the optical depth for each energy (Figure \ref{fig:optical depth}). In this context, we would like to mention  \cite{2019_Molkov}. They reported a cyclotron line at 10 keV along with it’s two harmonics at a particular pulse phase. Our study shows that we do not find any cyclotron line in the phase averaged spectra which is in agreement with their study and along with non-detection of any feature, it is customary to report the upper limit of the same.

Along with it we also look at the difference in the estimation of the strength of the surface magnetic field from spin-up rate and luminosity relation, and from the observed cyclotron line. We calculated the $\mu_{30}$ and $B_{12}$ from the proportionality factor k in the spin-up rate and luminosity relation in equation \ref{equ:nudot2} provided in \cite{2017_Sugizaki} for nine transient pulsars for which cyclotron line has been observed. Canonical values of $R_6=1$, $M_{1.4}=1$, $I_{45}=1$ for neutron stars have been assumed for the calculation.
We also calculated the magnetic field value from the observed cyclotron line energy using the relation $B_{12}= \frac{1}{\sqrt{1-x^{-1}}}\frac{E_{cyc}}{11.6keV}$ and compared them. Table \ref{B12} represents a large difference in these two $B_{12}$ values. The two estimates of magnetic field strength from torque-luminosity relation and cyclotron line energy differ by large factors, from $10^{-5}$ to 4.0. Part of these differences may arise if $R_{6}$, $M_{1.4}$, $I_{45}$, beaming factor, and effect of misalignment between spin axis and magnetic axis etc. for each individual source is different from 1.0. Another factor is the uncertainty in source distance and hence on luminosity. We note here the strong dependence of the magnetic moment on radius of the neutron star (third order) and distance (sixth order). A difference in distance by a factor of 1.5 can result in a factor of $\sim$10 difference in estimation of the magnetic field strength. Therefore, the actual surface magnetic field strength of GRO 2058+42 can be different from its estimated value from spin-up rate and luminosity relation.

\section*{Acknowledgements}
This research has made use of archival data and software provided by the High Energy Astrophysics Science Archive Research Center (HEASARC), which is a service of the Astrophysics Science Division at NASA/GSFC and the High Energy Astrophysics Division of the Smithsonian Astrophysical Observatory. We report the scientific results in this paper, based on data from the NuSTAR mission, a project led by the California Institute of Technology, managed by the Jet Propulsion Laboratory, and funded by the National Aeronautics and Space Administration.
For the purpose of analysis we have used the NUSTARDAS package which is developed by the ASI Science Data Center (ASDC, Italy) and the California  Institute of Technology (USA). We have also used the data from Fermi  Science  Support Center; long-term lightcurve from {\it{Swift}}-BAT transient monitor provided by the {\it{Swift}}-BAT team; and {\it{MAXI}}-GSC provided by RIKEN, JAXA and the MAXI team. 
SK thanks Saikat Das, who is a research fellow at RRI, for his help and suggestions.










\bibliography{bibtex}
\bibliographystyle{mnras}

\bsp	
\label{lastpage}
\end{document}